# Structure stability and magnetism in graphene impurity complexes with embedded V and Nb atoms

Jyoti Thakur[1], Manish K. Kashyap[1, †], Hardev S. Saini[2] and Ali H. Reshak[3,4]

[1]Department of Physics, Kurukshetra University, Kurukshetra-136119 (Haryana), India

[2]Department of Physics, Panjab University, Chandigarh-160014, India

[3]New Technologies - Research Centre, University of West Bohemia, Univerzitni 8, 306 14 Pilsen, Czech Republic

[4] Center of Excellence Geopolymer and Green Technology, School of Material Engineering, University Malaysia Perlis, 01007 Kangar, Perlis, Malaysia

[†]Corresponding Author's E-mail:- manishdft@gmail.com, mkumar@kuk.ac.in

## ABSTRACT

First-principles density functional theory (DFT) study of embedding V and Nb atom in monovacant and divacant graphene is reported. Complete/almost complete spin polarization is verified for V/Nb embedding in MV/DV graphene. The origin of magnetism has been identified via interaction of *3d*-states of embedded trnasition metal atom with *p*-states of inequivalent C atoms present in the vicinity of embedding site. Band structure analysis has been performed to address the semiconducting behavior of graphene in minority spin channel on embedding V/Nb atom. The isosurface plots also confirm the magnetic nature of present nanosystems. Our results reveal that these nanosystems have potential for futuristic applications such as spintronics, energy resources and high frequency transistors.

# 1. Introduction

A truly two-dimensional honeycomb structure of carbons atoms is famous as "Graphene", which can be exfoliated from graphite in the form of one-atom thick monolayer [1]. Right from its discovery, graphene is considered as excellent candidate for several fundamental and technological interests in view of various properties such as scalability, chemical stability, ballistic transport at room temperature [2-4], high mobility of charge carriers [5] and high elasticity [6]. Additionally, a weak spin-orbit interaction among carbon atoms leads to a long spin relaxation length on the surface of graphene [7]. Hence, graphene is a potential substrate to use in spintronic devices, recording media, magnetic inks, spin qubits, spin valves and nanomagnetic magnetism [8-9].

However, in spite of huge varieties of its novel applications, the use of graphene is rather limited due to its zero band gap [10-11]. Presently, many research efforts are directed towards inducing and fine tuning of band gap in graphene [12-13] and developing various approaches to effectively induce and manipulate the magnetic states in it which are crucial for its use in nanoscale devices [14-16]. Regarding magnetism, a lot of attentions were focused on the production of magnetic carbon from different routes like chemical vapour deposition [17-18], ion bombardment [19-20], nanofoam [21], ion implantation [22] etc. Spin polarized density functional theory (DFT) calculations of magnetic properties of the vacancies and vacancy hydrogen complexes of graphite [23] showed that these defects lead to a macroscopic magnetic signal as also observed in experiments [24]. Firstly, Lehtinen et al [25] reported on vacancies in graphene nanosheets and analyzed the role of adatoms in diffusion over surfaces to generate a magnetic moment of ~ 0.5 $\mu_B$ on graphene.

Experimental [26-27] and theoretical [28-29] investigations demonstrated that the most preferred way to incorporate transition metals (TMs) impurities in graphene is to attach them to vacancies. Recently, Gan et al [30] using high resolution transmission electron microscopy (HRTEM) observed substitutional Au and Pt atoms in graphene nanosheets increases the potential of graphene in spintronic applications. Wang et al [31] also found that a two step process is an efficient way to dope graphene nanosheets; (1) first create vacancies by high energy atom/ion bombardment and (2) then to fill these vacancies with desired dopants. Robertson et al [32] were able to locally control the formation of mono and divacancies in graphene at predefined sites with a focused electron beam at 80 keV, later these vacancies can act as trap sites for mobile Fe atoms.

On theoretical front, electronic and structural distortions in graphene induced by carbon vacancies and boron doping were studied by Faccio et al [33] and their results indicated that a boron atom behaves quite differently when occupies atomic position near or far from vacancies. When it is far away/ near from a vacancy, it slightly, enhances/destroys the magnetism in resultant nanosystem. Modelling of monovacant graphene doped with O and B atoms was performed by Kaloni et al [34] and their calculations demonstrated that B doping of oxidized vacancies is a successful approach to induce extended π-band magnetism. By controlling the O and B concentrations, it is possible to tune the magnetic state. Krasheninnikov et al [35] explained the attractive interactions between transition-metal atom impurities and vacancies in graphene using first-principles study.

On literature review, it has been observed that the study of magnetic substitutional impurities of TMs in graphene is of prime importance for both theoretical and experimental point of view. Further, it is also crucial to understand the origin of magnetism due to atomic defects in graphene nanomaterials. In present work, we have created the mono- and di-vacancies in graphene nanosheets and filled the one vacancy by a transition metal (X= V/Nb) dopants. The resultant nanostructure is planned to be examined for possessing spin polarization and magnetism. It is also aimed to realize how the TM gets strongly bound to defected graphene nanosheet and to observe the possible hybridization of C-sp$^2$ and TM-s,d orbitals. Further, an exhaustive comparison of peculiar electronic properties and magnetism of V/Nb atom-vacancy complexes in case of monovacancies (MV) and divacancies (DV) is also presented. This type of study is of prime importance as the resultant structures can be easily fabricated by electron irradiation to create defects first and depositing TM thereafter.

## 2. Computational Details

The TM (X = V/Nb) impurity embedded in MV and DV graphene were simulated by constructing 5×5×1 supercell of graphene. A vacuum layer of 20 Å thickness in the *z*-direction was inserted to avoid the interlayer interaction due to periodic boundary conditions. We employed density functional theory (DFT) [36] based projected augmented wave (PAW) method as implemented in VASP [37-38] for the relaxation of the systems under investigation. The generalized gradient approximation (GGA) under Perdew-Burke-Ernzerhof parameterization was used to construct exchange-correlation (XC) potentials [39]. The PAW method is an all electron description and is used to describe the electron-ion description. The cutoff energy for plane waves was set to be 400 eV, a 9×9×1 Monkhorst-Pack grid was used for sampling of the Brillouin zone during geometrical optimization. Conjugate gradient (CG) algorithm [40] was selected to relax present systems. All the internal coordinates were allowed to be relaxed until calculated Hellmann Feymann force on each atom became less than 0.02 eV $\text{Å}^{-1}$. Several test calculations for higher supercells; 7×7×1 and 8×8×1 were also performed which gave essentially the same results.

After relaxation, we performed electronic structure calculations of relaxed systems using the full-potential linearized augmented plane-wave (FPLAPW) [41] method within GGA as implemented in WIEN2k package [42]. In FPLAPW calculations, the core states were treated fully relativistically, whereas for the valence states, a scalar relativistic approach was used. Additionally, valence electronic wavefunctions inside the Muffin-tin sphere were expanded up to $l_{max}$=10. The radii of the Muffin-tin sphere ($R_{MT}$) for various atoms were taken in the present calculations such as to ensure nearly touching spheres. The plane wave cut off parameters were decided by $R_{mt}k_{max}$ =7 (where $k_{max}$ is the largest wave vector of the basis set) and $G_{max}$ =12 a.u.$^{-1}$ for fourier expansion of potential in the interstitial region. The k-space integration was carried out using modified tetrahedron method [43] with a k-mesh of 11×11×1 for the high resolution and better convergence in the calculations.

## 3. Results and Discussion

To begin with desired calculations, the graphene containing MV/DV was created by using 5×5×1 supercell of pristine graphene and removing anyone/two adjacent C-atoms. Thereafter at one vacant site, foreign atom (X=V/Nb) was embedded and the structure so obtained was relaxed to find the actual atomic positions and possible bonding. It was found that on embedding V/Nb in graphene with MV, the hexagonal symmetry near the vacant site remains maintained and the resultant nanosystem shows only substitutional effect with no shape transformation. However, embedding in graphene containing DV leads to shape change to a pentagonal structure at vacant sites via reordering of the chemical bonds between foreign V/Nb atom and C-atoms. The final relaxed structures in both cases are depicted in Fig.1.

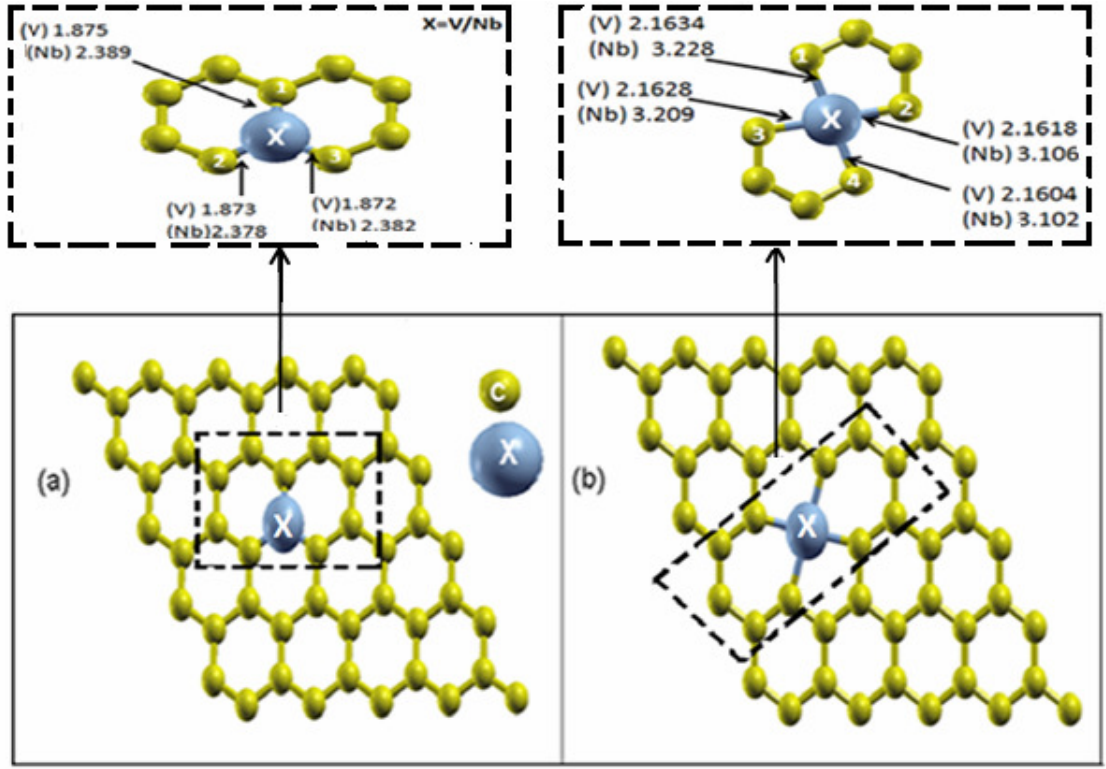

**Fig. 1**. Schematics for typical X=V/Nb atom embedding in graphene sheet containing (a) MV and (b) DV. Insets show the significant bond lengths of various C-X and C-C bonds in the vicinity of embedded site.

In order to estimate the stability of a structure containing embedded TMs (X=V/Nb) atom in graphene nanosheets containing MV /DV, the binding energies ($E_b^{MV+X}/E_b^{DV+X}$) were calculated as under:

$$E_b^{MV+X} = E_{tot}^{MV+X} - (E_{tot}^{MV} + E_{tot}^{X}) \quad (1)$$

$$E_b^{DV+X} = E_{tot}^{DV+X} - (E_{tot}^{DV} + E_{tot}^{X}) \quad (2)$$

where $E_{tot}^{MV+X}/E_{tot}^{DV+X}$ is the ground state energy of metal mono-/di-vacancy graphene complex and the bracketed term in these equations represents the sum of ground state energies of reconstructed naked mono/di-vacancy ; $E_{tot}^{MV}/E_{tot}^{DV}$ [25,44] and isolated TM (X) atom; $E_{tot}$. The computed binding energies are presented in Table-I. We observe that embedded V atom in graphene containing MV/DV are found to be most energetically stable

(Table-I) as compared to embedded Nb atom in it. Further, these findings are good agreements in previous works reported [45-46] in literature.

**Table-I:** Calculated bond lengths (Å) and binding energies (eV) of TM (X= V/Nb) atoms embedded in graphene sheet with (a) MV and (b) DV.

| Embedded TM (X=V/Nb) in graphene containing | $E_b$ (eV) | Bond length (Å) | |
|---|---|---|---|
| MV | -7.892 (V) | V-$C_1$ =1.876 | Nb-$C_1$=2.389 |
|  | -6.584 (Nb) | V-$C_2$ =1.873 | Nb-$C_2$=2.378 |
|  |  | V-$C_3$ =1.872 | Nb-$C_3$=2.382 |
| DV | -3.489(V) | V-$C_1$ =2.163 | Nb-$C_1$=3.228 |
|  | -2.769(Nb) | V-$C_2$ =2.162 | Nb-$C_2$=3.106 |
|  |  | V-$C_3$ =2.163 | Nb-$C_3$=3.209 |
|  |  | V-$C_4$ =2.160 | Nb-$C_4$ =3.102 |

In case of embedding X-atom in graphene containing MV/DV, three/four neighbouring C-atoms make bond with embedded atom. There is a slight difference in relaxed bond lengths of various bonds (mentioned in Table-I) in all 4 cases but these bond lengths are definitely smaller than the standard C-C bond length (2.46Å) of pristine graphene. Further, X-C bond length increases for both MV and DV cases, as we move from embedding of V to Nb atom due to increase in atomic size of latter.

Fig. 2 manifests the symmetrical DOS for pristine graphene in both spin channels which leave it in semimetallic state [47]. On embedding the X atom (X=V/Nb) in MV/DV case, the total DOS gets altered and localized levels start appearing, especially in the vicinity of $E_F$. Embedding V/Nb atom in MV case generates complete spin polarization however the same embedding in DV case creates large spin polarization, but not completely 100% (Table-I/II). However, this amount of polarization is also appreciable as it is capable of transporting spin polarized currents in spin quantum information devices and spin filter devices as like 100% spin polarization. The localized levels appearing in all embedded cases are the consequence of reformation of energy bands due to interaction of C-s and C-p states with d-states of foreign V/Nb atom.

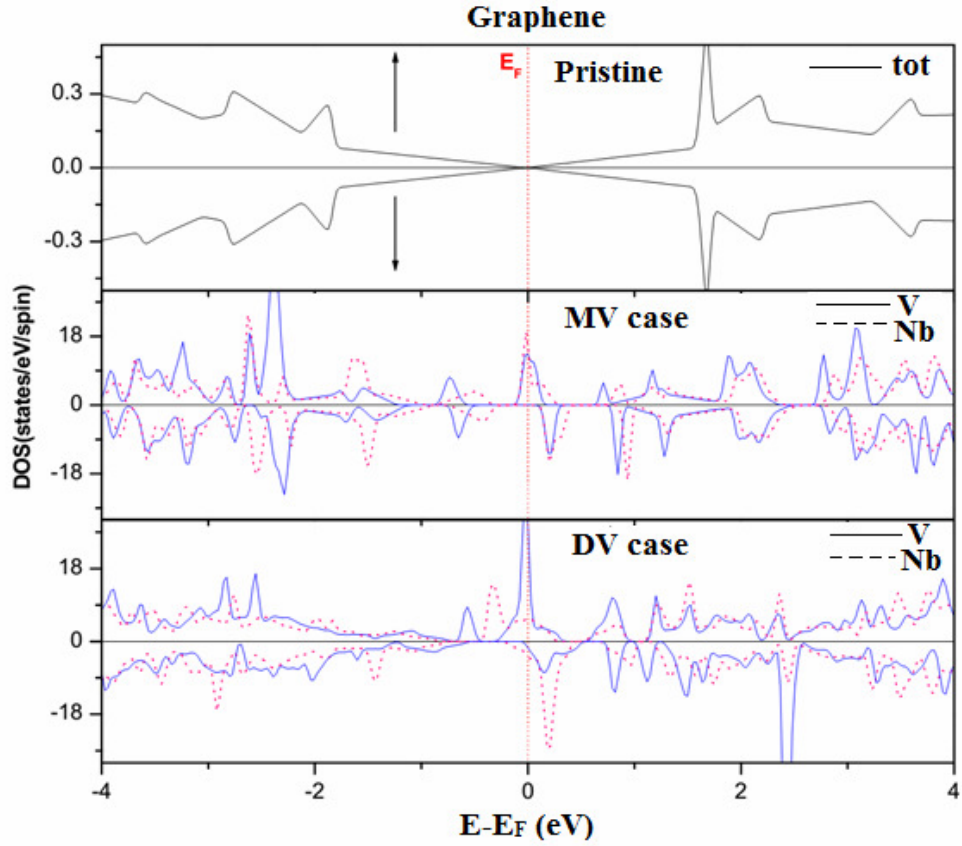

**Fig.2.** Calculated spin polarized total density of states (TDOS) of pristine graphene, and X-embedded graphene (X= V/Nb) containing MV/DV.

To explore the contribution of various states in total DOS of resultant nanosystems, V-embedded graphene containing MV and DV were investigated (Fig. 3). It is found that V-d states and nearest neighboring C-p states are located at $E_F$ whereas in minority spin, these states are almost absent. This presents a direct evidence of dependence of spin polarization on interactions of V-d and C-p states. The C atoms away from embedded site ($C_{far}$) represent almost identical DOS in the vicinity of $E_F$, confirming that they have no role to decide any spin conduction. Nb-embedding in graphene containing MV/DV (not shown for brevity) can also be explained on the similar lines.

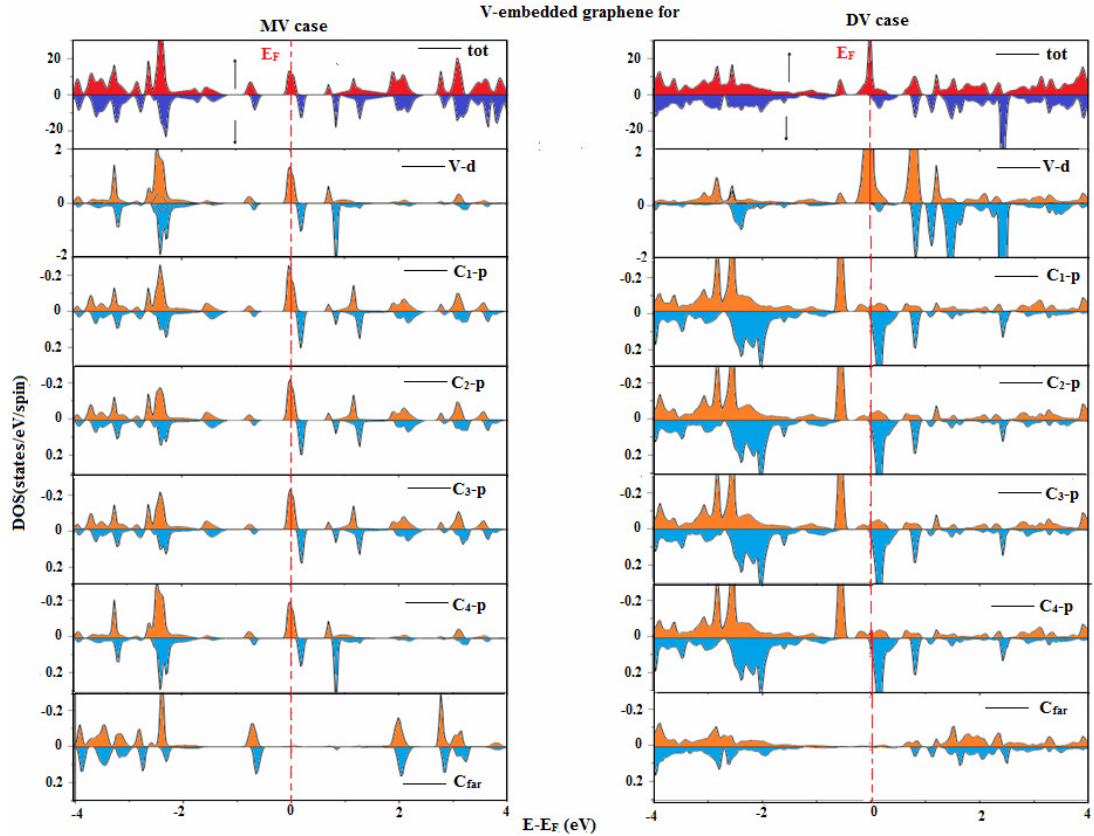

**Fig. 3**. Calculated spin resolved partial density of states (PDOS) of V-embedded graphene with MV/DV

Fig. 4 manifests the bandstructure of V-doped graphene included MV along with its total DOS. The bandstructure also confirms the prediction of half metallicity in minority spin channel as governed by DOS plots. The present nanosystems contain direct bandgap along the K-K direction. In majority spin channel (MAC), the Dirac cone of graphene along symmetry point (K) at $E_F$ undergoes red shift and its complete part remains present in valence band now after embedding V-atom. The upper band of the cone gets flattened as well due to the hybridization of V-d states and C-p states. However, in minority spin channel (MIC), the upper band of Dirac cone remains fixed at $E_F$ as in case of pristine graphene but the lower band shows the red shift due to hybridization of V-d and C-p states. This opens up the band gap of graphene in MIC. Further, large number of occupation states is available in conduction band which can be easily populated via easy excitation of electrons. Thus, the semiconducting properties can be drawn from MIC of graphene which enhances its usefulness in nano-opto electronic devices. The perturbation of Dirac cone was also observed previously in F-, Ge-, Li- and K-intercalated graphene on SiC [48-50].

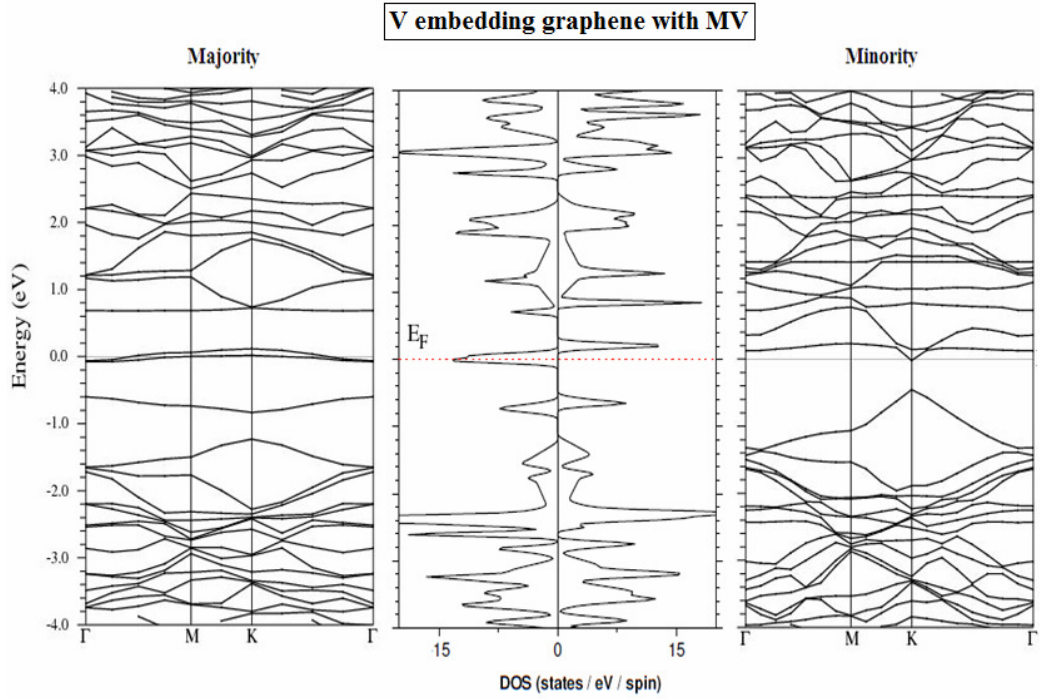

**Fig.4.** Calculated spin polarized bandstructure along with total DOS of embedded V atom in graphene containing MV

Pristine graphene has zero magnetic moment; but in X (X=V/Nb) embedded graphene nanosystems with MV/DV, C atoms in the vicinity of vacant site become non-equivalent. The X atom and non-equivalent C atoms containing spin polarized DOS induce the magnetic moment on graphene (Table-II). The total moment is proportional to net difference of DOS of electrons in two spin channels. Therefore, an appreciable moment of 1.33$\mu_B$/ 3.35 $\mu_B$ has been found in V embedded for MV/DV case. The corresponding moment in case of Nb embedding in MV/DV case is somewhat smaller than in V-embedding. The other C atoms, which are away from embedded site, do not possess any significant magnetic moments. However, some moment is found to be distributed evenly in the interstitial regions among C atoms, which is also the case for $g$-$B_3N_3C$ nanosheet; graphite like material [51]. This indicates that the magnetic moment distributed over the interstitial regions is crucial and should be considered for counting the total magnetic moment. We noticed that present TM complexes in graphene with MV are magnetic due to partial filled TM-d states. The magnetic behaviour of TMs (X=V/Nb) complexes in DV case is even more interesting.

It is observed that X embedding in graphene with DV results in a ''cross'' configuration (Fig. 1). A larger ''hole'' at the di-vacancy site produces weaker interactions of the embedded atom with the ligand bonds which yields higher spin states of the complex. V ($3d^3\ 4s^2$) atom has 5 valence electrons. Out of these, three valence electrons contribute to V-C σ covalent bonds. The fourth electron plays the same role as by that of missing C-atom at vacant site. This results in formation of one π-bond, out of the plane of graphene. The additional electron left behind is free and can enter easily in non-bonding orbital. Due to this electron, additional magnetic moment appears on embedded graphene containing MV. The moment of Nb-embedded graphene containing MV decreases slightly due to presence of radial node of the Nb-4d orbital. A weaker p-d hybridization due to this node results in decrease of magnetic moment as compared V-embedded graphene with MV. On applying similar logic to V-embedding in DV case, we found that the foreign V atoms makes four local σ bonds with nearby C atoms (Fig. 2). Still one extra electron remains unsaturated on V-atom. Further, the V-C bond length is large in this case which enhances localization of the additional electron. The formation of σ bonds and more localization increases the availability of the more states for free electrons due to which the magnetic moment enhances up to $3.00\mu_B$. The decrease in magnetic moment for Nb embedded graphene containing DV can be justified on simlar lines as in the case of MV graphene (Table-II)

**Table-II: Calculated spin polarization P(%), magnetic moment M ($\mu_B$) of X (X= V/Nb) embedded in graphene with MV/DV, magnetic moment appears on individual V/Nb atom, magnetic moment on first/second nearest carbon atoms ($M_{C1}$/ $M_{C2}$) and interstitial regions ($M_{inter}$)**

| Graphene | P(%) | M ($\mu_B$) | $M_{V/Nb}$ | $M_{C1}$ | $M_{C2}$ | $M_{inter}$ |
|---|---|---|---|---|---|---|
| MV Graphene | V= 98 | 1.35 | 0.689 | 0.028 | 0.027 | 0.121 |
|  | Nb= 96 | 0.99 | 0.737 | 0.035 | 0.034 | 0.033 |
| DV Graphene | V=100 | 3.35 | 1.866 | 0.118 | 0.116 | 0.832 |
|  | Nb=98.6 | 2.30 | 1.095 | 0.114 | 0.112 | 0.228 |

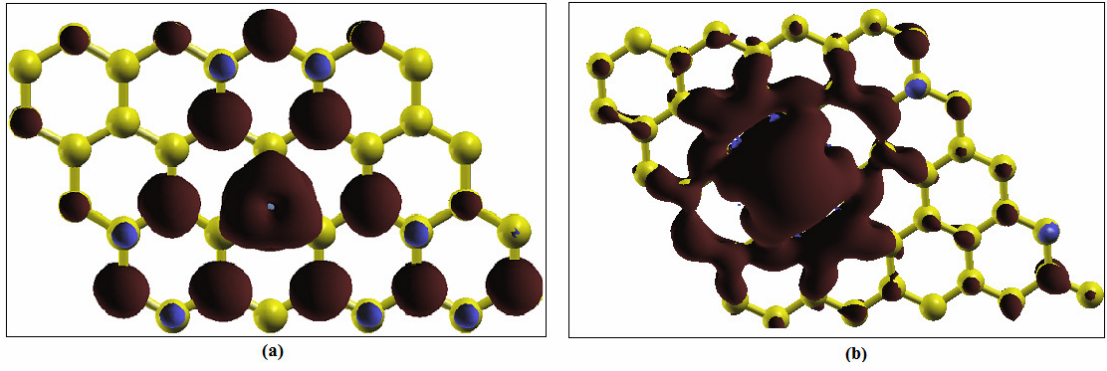

**Fig. 5**. Spin density difference isosurfaces for X (X=V/Nb) embedded graphene for (a) MV and (b)DV cases. The isosurfaces correspond to a value of $3 \times 10^{-3}$ electrons/Å$^3$. Brown/blue color represnts positive/negative values of accumulated spin density.

The isosurface analysis is also performed to explain the magnetic ordering in the present nanosystems. To analyse the same, we have represented spin density difference isosurfaces $\Delta\rho(r)$ for V embedding in graphene with (a) MV and (b) DV in Fig. 5, where

$$\Delta\rho(r) = \rho_{MAC}(r) - \rho_{MIC}(r) \qquad (3)$$

The positive value of $\Delta\rho(r)$ indicates the existence of magnetic moment in a specific region near the vacancy in both MV and DV cases. This results in stable ferromagnetic state for both cases. Further, $\Delta\rho(r)$ is more positive in DV case as due more localization, therefore, the availability of free electrons in the vicinity of $E_F$ increases. This governs the enlarged moment for DV case. For Nb embedding, similar lines may be quoted. Qualitatively, our findings agree well with previous reports for 3d transition metal atoms decorated graphene [52-55].

## 4. Summary and conclusions:

The electronic and magnetic properties of X embedded (X = V and Nb) graphene with MV/DV have been studied using density functional theory (DFT) approach. The PAW method within VASP was used to predict the structure stability and to find final relaxed structure for present nanosystems. Our FPLAPW calculations show that the electronic properties and magnetic response of graphene containing MV/DV get modified drastically. The bandstructures of present nanosystems have been explained to identify the shifting of Dirac cone of graphene and hence to justify the semiconducting behaviour of graphene in minority spin channel. The isosurfaces plots confirm the stability of ferromagnetic state of graphene on embedding V/Nb atoms. The magnetic moment is mainly contributed by doped TM atom with appreciable contributions from nearby C atoms as well. A complete spin polarization (100%) in MV case and nearby 100% spin polarization in DV case make resultant nanosytems most probable nanomaterial for the development of nanoscale devices for spintronic applications, spin filters and injector devices. Since the presence of transition metal impurities could be detected easily by scanning tunnelling microscopy (STM), high resolution travelling electron microscopy (HRTEM) and X-ray absorption technique, thus the present investigations leave a definite experimental scope to work with the nanosystems under investigation.

**Acknowledgement**

The computational facilities used for the work are supported by UGC, INDIA, grant No. 41-922/2012 (SR) sanctioned to M. K. Kashyap. H.S. Saini acknowledges UGC, INDIA for providing financial support as Dr. D. S. Kothari post doctoral fellowship. For the author – A. H. Reshak, the result was developed within the CENTEM project, reg. no. CZ.1.05/2.1.00/03.0088, cofunded by the ERDF as part of the Ministry of Education, Youth and Sports OP RDI programme and, in the follow-up sustainability stage, supported through CENTEM PLUS (LO1402) by financial means from the Ministry of Education, Youth and Sports under the "National Sustainability Programme I. Computational resources were provided by MetaCentrum (LM2010005) and CERIT-SC (CZ.1.05/3.2.00/08.0144) infrastructures".